\documentclass{aa}

\usepackage{txfonts}
\usepackage[dvips]{graphicx}
\usepackage{longtable}
\usepackage{natbib}
\bibpunct{(}{)}{;}{a}{}{,}

\begin{document}

\title{Blazar sequence -- an artefact of Doppler boosting}

\author{E. Nieppola\inst{1}, E. Valtaoja\inst{2,3}, M. Tornikoski\inst{1}, T. Hovatta\inst{1}, M. Kotiranta\inst{1}  
       }

\institute{Mets\"ahovi Radio Observatory, TKK, Helsinki University of Technology, Mets\"ahovintie 114, 
           FIN--02540 Kylm\"al\"a, Finland\\
		\email{elina.nieppola@tkk.fi} 
	\and
	Tuorla Observatory, V\"ais\"al\"antie 20, 21500 Piikki\"o, Finland
	\and
	Dept. of Physical Sciences, University of Turku, 20100 Turku, Finland\\
}

\date{Received / Accepted}


\abstract{The blazar sequence is a scenario in which the bolometric luminosity of the blazar governs the appearance of its spectral energy distribution. The most prominent result is the significant negative correlation between the synchrotron peak frequencies and the synchrotron peak luminosities of the blazar population.}{Observational studies of the blazar sequence have, in general, neglected the effect of Doppler boosting. We study the dependence of both the synchrotron peak frequency and luminosity with Doppler-corrected quantities.}{We determine the spectral energy distributions of 135 radio-bright AGN and find the best-fit parabolic function for the distribution to quantify their synchrotron emission. The corresponding measurements of synchrotron peak luminosities and frequencies are Doppler-corrected with a new set of Doppler factors calculated from variability data. The relevant correlations for the blazar sequence are determined for these intrinsic quantities.}{The Doppler factor depends strongly on the synchrotron peak frequency, the lower energy sources being more boosted. Applying the Doppler correction to the peak frequencies and luminosities annuls the negative correlation between the two quantities, which becomes \textit{positive}. For BL Lacertae objects, the positive correlation is particularly strong.}{The blazar sequence, when defined as the anticorrelation between the peak frequency and luminosity of the synchrotron component of the spectral energy distribution, disappears when the intrinsic, Doppler-corrected values are used. It is an observational phenomenon created by variable Doppler boosting across the synchrotron peak frequency range.}


\authorrunning{Nieppola et al.}
\titlerunning{Blazar sequence and Doppler boosting}

\maketitle


\section{Introduction}

The most powerful active galactic nuclei (AGN) are active emitters at almost all frequencies and exhibit both strong flaring and typically high polarization. These sources are referred to, in general, as blazars. The source of their power is a supermassive black hole embedded in the host galaxy, which generates a massive accretion process. The bulk of the continuum radiation from these sources does not, however, originate in the nucleus, but in two jets that emerge symmetrically from the core. A tell-tale sign of the presence of jets is strong radio continuum, which is believed to be produced via synchrotron process by relativistic electrons spiralling in the magnetic field of the jet. This synchrotron radiation can be seen in the blazar spectral energy distribution (SED) as a bump at radio to X-ray frequencies in the log $\nu$ -- log $(\nu F_{\nu})$ -representation. It is followed by another bump, which is often attributed to the inverse Compton process.

The shape and position of the synchrotron SED component in the log $(\nu F_{\nu})$ versus log $\nu$ coordinate system and the mechanisms that control its properties have been discussed for a number of years. It was first suggested by \citet{padovani95_I} that the peak of the synchrotron component can occur at any frequency from radio to soft X-rays, producing detectable differences in the source properties, such as broad band spectral indices. The position of the synchrotron component peak on the log $\nu$ -axis, the synchrotron peak frequency $\nu_p$, was linked to the intrinsic synchrotron peak luminosity, $\nu_p L_{\nu,p}$, by \citet{fossati98} and \citet{ghisellini98}. They proposed that there was a negative correlation between the two quantities, that is sources with their peak frequency in the radio band would be more luminous than those that had their peak in the X-ray band. This correlation would be created by decreasing intrinsic power and decreasing cooling effects, caused by external radiation field, as the synchrotron peak moves from low to high energies \citep{ghisellini98}. The blazar luminosity sequence has been studied by many authors \citep[e.g.,][ for a concise review of the present status of the blazar sequence, see \citet{padovani07}]{padovani03, caccianiga04, anton05}. Lately, evidence against it has been mounting; as the number of blazars known has increased, it has been found that the previously-observed correlation does not describe the newer data sets. \citet{ghisellini08} attempted to explain these outliers using a revised theoretical blazar sequence, which related the SED shape to the mass of the central black hole and its accretion rate. This new scheme reduced the tightness of the correlation between the peak frequency and luminosity, allowing the presence of high-luminosity high-$\nu_p$ objects, and even predicting the existence of low-power low-$\nu_p$ objects. However, the negative trend between the two quantities was, in general, expected to hold. In \citet[][hereafter Paper I]{nieppola06} the synchrotron peak frequencies of more than 300 BL Lacertae objects (BL Lacs, BLOs) were determined and used to plot the $\nu_p L_{\nu,p}$ versus $\nu_p$ -correlation. No significant anti-correlations could be found using this extensive sample, which included many high-energy sources.

One important matter overlooked in these observational studies of the blazar sequence was the effect of Doppler boosting. When the source of the continuum radiation (i.e., the shocked plasma in the jet) is moving towards us, the observed radiation is blueshifted and enhanced significantly \citep{blandford79}. The amount of Doppler boosting depends on the viewing angle of the jet, $\theta$, and the Lorentz factor $\gamma$. It is therefore clear that the Doppler factor $D$ is by no means similar for all AGN, or even for all blazars, which implies that it can significantly effect the appearance of the blazar sequence. In the original blazar sequence papers \citep{fossati98, ghisellini98} and in theoretical work \citep{ghisellini08}, the Doppler factor was assumed to be constant irrespective of the peak frequency $\nu_p$. In this paper, we demonstrate that this assumption is inaccurate. We examine the $\nu_p$ versus $\nu_p L_{\nu,p}$ -correlation for a complete sample of AGN using Doppler-corrected values based on new estimates of Doppler-boosting factors (Hovatta et al. 2008, in preparation). In this way, we are able to determine if there is any pattern in the intrinsic, true properties of the sources. 

The paper is organized as follows: in Sect.~\ref{sam} we present the source data that we study, and in Sect.~\ref{methods} we describe the methods of acquiring the data. In Sect.~\ref{results}, we present the results, and finish with a discussion and conclusions in Sects.~\ref{dis} and \ref{con}, respectively. Henceforth, all luminosities are frequency dependent. For clarity, we denote luminosities simply with $L$ instead of $L_{\nu}$, and use the subscript $p$ to denote peak frequency and luminosity of the synchrotron component determined directly from the SED. Throughout this paper, we assume $H_{0}=65\,\textrm{km}\,\textrm{s}^{-1}\,\textrm{Mpc}^{-1}$ and $\Omega_0=1$.

\section{Sample}
\label{sam}

We study sources selected from the Mets\"ahovi Radio Observatory source list, which meet one or both of the following criteria:
\begin{enumerate} 
\item The source has a declination of $\geq 0\deg$ and the highest observed flux exceeds 1.2 Jy at 37 GHz.
\item The source has a Doppler factor determined by either \citet{lahteenmaki99_tfdIII} or Hovatta et al. 2008 (in preparation).
\end{enumerate}

Sources meeting the first criterion comprise a complete northern flux-limited radio-loud AGN sample; they constitute a large majority of the entire sample, 113 sources out of 135. The remaining sources are radio loud as well, but do not reach the flux limit of 1.2 Jy, or are southern sources. They are included because they have well sampled radio flares at 22 or 37 GHz, allowing a reliable determination of the Doppler factor from variability data. The full sample consists of 34 highly polarized quasars (HPQ), 33 quasars of low polarization (LPQ), 31 BL Lacertae objects (BLO), 26 quasars with no polarization data (QSO), 9 radio galaxies (GAL), and 2 unclassified objects. HPQs have a measured polarization of $\geq 3$ \% in the literature and LPQs have a polarzation of $< 3$ \%. BLOs are included in \citet{veron00}, or defined as such in the literature. GALs are radio galaxies of varying properties, which all are non-quasars.

The full sample is listed in Table~\ref{sample}. Columns (1) and (2) provide alternative names for the source and Cols. (3) and (4) indicate the right ascension and declination, respectively. Column (5) indicates whether the source belongs to the 1.2 Jy complete sample. Column (6) provides the classification of the source, our measeurement of the synchrotron peak frequency (see Sect.~\ref{seds}) is listed in Col. (7), and Col. (8) gives the redshift of the source. Finally, Col. (9) indicates the sources that belong to the blazar sample discussed in Sect.~\ref{peak_corr}. 
\addtocounter{table}{1}

\section{Methods}
\label{methods}
\subsection{Spectral energy distributions}\label{seds}

The spectral energy distributions of our selected sample were determined in a similar manner to those of Paper I. A significant quantity of multi-frequency data were collected from various sources, including the CATS database\footnote{http://cats.sao.ru} of the Special Astrophysical Observatory \citep{verkhodanov05} and the NED database\footnote{http://nedwww.ipac.caltech.edu/} at IPAC. Most sources have a low-energy synchrotron peak and we were able to find a substantial amount of radio data. The majority of the high frequency radio data were acquired from Mets\"ahovi Radio Observatory \citep[][and some unpublished data]{salonen87, terasranta92, terasranta98, terasranta04, terasranta05, nieppola07}. IR-data from IRAS or 2MASS catalogues, or from ISO \citep{padovani06} were also available for most sources in the aforementioned databases. In addition to X-ray datapoints found in the databases, we used 1 keV monochromatic fluxes from \citet{donato01}. 

The SEDs were plotted in log $\nu$ -- log $(\nu F_{\nu})$ parameter space and are presented in Fig.~\ref{sedfig} (published electronically). The synchrotron components of the SEDs were fitted by a simple parabolic function \begin{equation}\label{parabola}\textrm{log}\,(\nu F_{\nu}) = A\,(\textrm{log}\,\nu)^2+B\,(\textrm{log}\,\nu)+C,\end{equation} where A, B, and C are constants. Thus, the synchrotron peak frequency is given by \begin{equation}\textrm{log}\,\nu_{p} = \frac{-B}{2A}.\end{equation} With knowledge of $\nu_p$, we can calculate the peak flux density, $\nu_p F_{\nu, p}$, using Eq.~\ref{parabola}. The peak frequencies obtained in this work are listed in Col. (7) of Table~\ref{sample}.

In Fig.~\ref{sedfig}, datapoints included in the fitting are marked by filled circles and other datapoints are marked by open rectangles. For the sake of consistency, we excluded datapoints below 1 GHz from all fits. Below that frequency, the SEDs often feature an additional flattening originating from an extended component; we therefore applied a cutoff to avoid any contamination in the jet synchrotron fit. The synchrotron peak frequency could be determined for all but two sources: 1019+222 and 1845+797. These objects have peculiar spectral shapes which could not be fitted accurately using a parabolic function.

Our sample consists of radio-bright AGN, which typically have a relatively low synchrotron peak frequency. Thus, in most SED fits, we did not include the X-ray datapoints in the synchrotron component. Our decision was made based on a careful visual inspection and the X-ray spectral index when possible. In some powerful BLOs, such as 0219+428 and 1652+398, the synchrotron component extends to the X-ray domain. In the case of 0735+178, the X-ray data are also included in the synchrotron component in the interest of constraining the fit and providing a more accurate estimate of $\nu_p$. The X-ray indices of the SEDs are taken from \citet{donato01}. For clarity, they are plotted in the energy range 1 - 10 keV and are normalized with respect to the 1 keV monochromatic flux listed in \citet{donato01}.

There was one case, namely 0430+052, where the synchrotron peak was exceptionally well sampled, but the parabolic function did not fit the data very well. To avoid using an obviously flawed $\nu_p$, we assigned a new value using the true peak defined by the highest point of the synchrotron bump. Thus, we obtained log $\nu_p$ = 13.95, instead of log $\nu_p$ = 15.94 provided by the fit. The same approach was considered for 0316+413, but the two methods yielded almost identical results. In the analyses, we used the fit value log $\nu_p$ = 14.40.

\subsection{Doppler factors}\label{exp_fits}

To calculate the de-boosted, intrinsic values of $\nu_p$ and $\nu_p L_p$, we need to calculate the Doppler boosting factors $D$. These have been calculated in the past using several methods \citep[e.g.][]{ghisellini93, guijosa96, guerra97, lahteenmaki99_tfdIII, jorstad05, wu07}. In this work, we use the $D_{var}$ -values of Hovatta et al. 2008 (in preparation), which have been determined from total flux density variability data in the same manner as in \citet{lahteenmaki99_tfdIII}. The flux curves from long-term monitoring at 22 and 37 GHz at Mets\"ahovi Radio Observatory are decomposed into exponential flares. The exponential fit provides the necessary parameters for the calculation of the variability brightness temperature $T_{b, var}$; comparing $T_{b, var}$ with the equipartition value $T_{eq}$ then provides the amount of boosting. Both \citet{lahteenmaki99_tfdIII} and Hovatta et al. (2008) used the equipartition value of $5 \times 10^{10}$ K proposed by \citet{readhead94}. 

We were able to derive $D_{var}$ for 89 sources in our sample. The remainder of the sample had to be excluded from the study of intrinsic properties. The values of $D_{var}$  and a detailed explanation of their determination will be published in a future paper (Hovatta et al. 2008).

\section{Results}
\label{results}

\subsection{The dependence of $D_{var}$ and $\nu_p$}
\label{d_vs_nu}

\begin{figure}
\resizebox{\hsize}{!}{\includegraphics{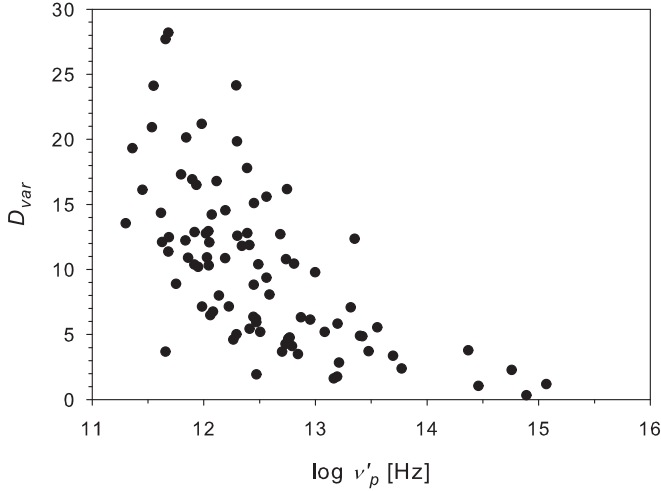}}
\caption{The variability Doppler factors $D_{var}$ plotted against the synchrotron peak frequency. The $\nu'_p$ -values are $D$-corrected.}
\label{Dvar_vs_peak}
\end{figure}

For studies of the blazar sequence, the crucial question is whether $D$ correlates with $\nu_p$ and thus changes the shape of the redshift-corrected $\nu_p L_p$ versus $\nu_p$ sequence, which has been considered in previous papers.
We therefore checked how measurements of $D_{var}$ from Hovatta et al. 2008 correlate with the $D$-corrected $\nu_p$. The Doppler correction of peak frequency was performed using equation \begin{equation}\label{dopfreq}\nu'_p=\nu_p\left(\frac{1+z}{D_{var}}\right),\end{equation} where the primed symbol represents the parameter value in the source rest frame. The result is presented in Fig.~\ref{Dvar_vs_peak}. The negative correlation is obvious, and the Spearman rank correlation test provides a correlation coefficient $\rho$ = -0.698 and a probability $P <$ 0.001 of no correlation. The correlation remains similar if we use values of $\nu_p$ uncorrected for the Doppler factor. The sources with their synchrotron peak at low energies are significantly more boosted than high-$\nu_p$ sources. A similar negative correlation ($\rho$ = -0.654 and $P$ = 0.004) can be obtained using the Doppler factors of \citet{jorstad05}, determined from variability data. We measured this also using the Doppler factors of \citet{ghisellini93}, calculated from the difference between the observed X-ray flux and one predicted by synchrotron self-Compton theory, based on the observed VLBI data. In this case, the correlation is very strong ($\rho$ = -0.517 and $P <$ 0.001). Therefore, we conclude that dependence is similar irrespective of the method used to calculate the $D$ -factors, and the negative correlation between $D_{var}$ and $\nu_p$ is robust.

Figure~\ref{Dvar_vs_peak} also depicts the broadness of the range of $D_{var}$, stretching from close to 0 to 30. At low $\nu_p$ in particular, it is crucial to have an accurate measurement of the source Doppler factor to obtain credible results of the intrinsic properties. \citet{wu07} determined Doppler factors derived from the relation between the observed 5 GHz core luminosity and a theoretical one, calculated from the total radio power at 408 MHz. Using log $\nu_p$ -values from Paper I, they also plotted the dependence between $D$ and $\nu_p$. Their Fig.~13 bears a striking resemblance to Fig.~\ref{Dvar_vs_peak} of this work.
 
\subsection{$\nu_p L_p$ vs. $\nu_p$ -correlation}
\label{peak_corr}

\begin{figure}
\resizebox{\hsize}{!}{\includegraphics{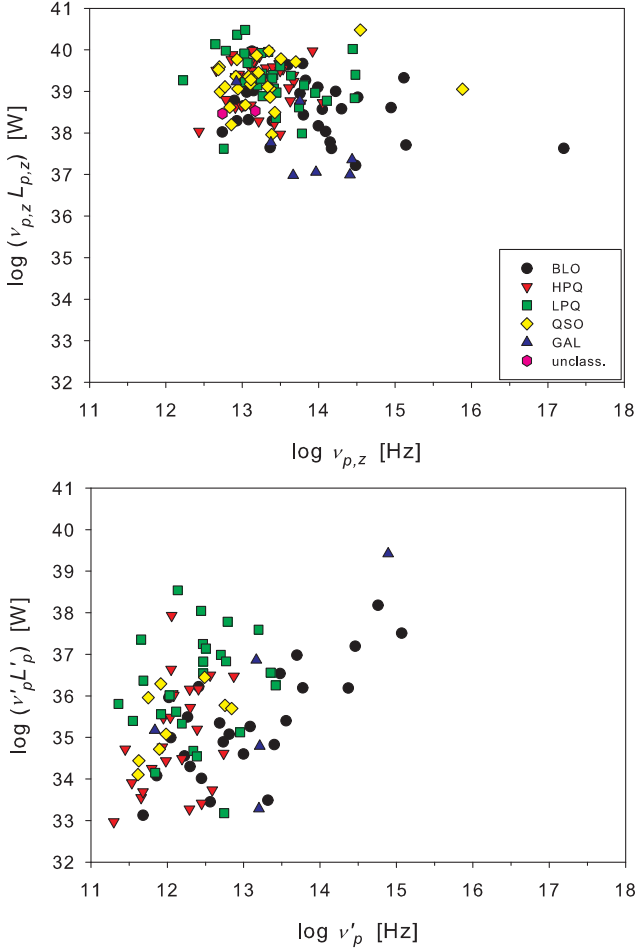}}
\caption{The peak luminosity of the synchrotron component plotted versus the synchrotron peak frequency. In the top panel, the values are $z$-corrected, and in the bottom panel Doppler-corrected. AGN classes are marked with different symbols (see legend).}
\label{lum_vs_peak_z}
\end{figure}

We examined the $\nu_p L_p$ versus $\nu_p$ -correlation using first redshift-corrected values and then Doppler-corrected, intrinsic values. We observe the results in Fig.~\ref{lum_vs_peak_z}. In the top panel, we display the observational, $z$-corrected correlation, and in the bottom panel the intrinsic, $D$-corrected correlation. The frequency redshift-correction is completed using the equation \begin{equation}\label{zcorr}\nu_{p,z}=\nu_p(1+z),\end{equation} and the redshift-corrected luminosity $L_{p,z}$ is calculated with the equation \begin{equation}\label{luminosity}L_{p,z}=\frac{4 \pi d_L^2}{1+z} F_{\nu,p},\end{equation} where $d_L$ denotes luminosity distance and $F_{\nu,p}$ is determined from the SED. The Doppler-corrections are performed using Eq.~\ref{dopfreq} and the equation \begin{equation}\label{doplum}L'_p=L_{p,z}\left(\frac{1+z}{D_{var}}\right)^{3+\alpha},\quad\mbox{assuming $F \propto \nu^{-\alpha}$.}\end{equation} In all equations, the subscript $z$ represents redshift-corrected values and primed values are both redshift and Doppler-corrected. When using Eq.~\ref{doplum}, we assumed a spectral index $\alpha = 1$ according to the definition of the synchrotron peak of the SED. All equations concerning redshift- and Doppler-corrections can be found in e.g. \citet{kembhavi99}.

The blazar sequence is represented by a negative correlation between the synchrotron peak frequency $\nu_p$ and peak luminosity $\nu_p L_p$. When our observational, redshift-corrected correlation is considered, we find that a significant correlation is also measured for our sample. The Spearman rank correlation test measures $\rho$ = -0.248 and $P$ = 0.002 for the entire sample. The Spearman test results for different AGN subclasses can be seen in Table~\ref{spearman_zcorr}: only the correlation of LPQs is found to be significant. Considering the entire sample, the trends visible in the top panel of Fig.~\ref{lum_vs_peak_z} are similar to those of the blazar sequence scenario found in several previous papers.

The full sample of our study includes many sources that are not traditionally classified as blazars, although the blazar classification is not clearly defined. For consistency, we identified the bona fide blazars in our sample and measured the correlation for their data alone. We classified a source as a blazar if it is included in one of the following: i) the total blazar sample of \citet{fossati98}, ii) the extragalactic radio sources list of \citet{wall85} and has $\alpha_{\textrm{\tiny{2.7-5 GHz}}} \leq$ 0.5 ($S \propto \nu^{-\alpha}$), iii) the DXRBS blazar sample of \citet{perlman98}; or is classified as a BLO according to Table~\ref{sample}. There were 65 blazars that fulfilled at least one of these criteria, and they are marked in Col. (9) of Table~\ref{sample}. From Table~\ref{spearman_zcorr}, we observe that the blazar sequence can be found in our data when observational quantities are used, because there is a significant anticorrelation between the $z$-corrected $\nu_p L_p$ and $\nu_p$ for blazars alone. We list results also for the complete 1.2 Jy AGN subsample and separately for blazars belonging to the 1.2 Jy sample. For both subsets of data separately, the negative correlation is clearly significant.

\begin{table*}
\caption{Spearman rank correlation test results for the whole sample and AGN subclasses for the log $\nu_p L_p$ vs. log $\nu_p$ -correlation. On the left side is the $z$-corrected correlation, and on the right side the $D$-corrected one. Galaxies (GAL) have been omitted due to small sample size.}
\label{spearman_zcorr}
\centering
\begin{tabular}{c c c | c c c}
\hline\hline
\noalign{\smallskip}
\multicolumn{3}{c}{$z$-corrected} & \multicolumn{3}{c}{$D$-corrected}\\
Class & $\rho$ & $P$ & Class & $\rho$ & $P$ \\
\hline
All & -0.248 & 0.002 & All & 0.353 & $<$ 0.001\\
BLO & -0.220 & 0.126 & BLO & 0.642 & $<$ 0.001\\
HPQ & -0.049 & 0.390 & HPQ & 0.354 & 0.041\\
LPQ & -0.362 & 0.019 & LPQ & 0.236 & 0.128\\
QSO & 0.110 & 0.296 & QSO & 0.550 & 0.063\\
Blazars & -0.329 & 0.004 & Blazars & 0.366 & 0.003\\
1.2 Jy AGN sample & -0.215 & 0.013 & 1.2 Jy AGN sample & 0.340 & 0.002\\
1.2 Jy blazars & -0.287 & 0.020 & 1.2 Jy blazars & 0.353 & 0.009\\
\hline
\end{tabular}
\end{table*}

We note that the observational anticorrelation detected here is not in disagreement with the conclusion of Paper I, which reported a lack of correlation between $\nu_p L_p$ and $\nu_p$ for BLOs. The range of $\nu_p$ was very extended in Fig.~6 of that paper, whereas in Fig.~\ref{lum_vs_peak_z} here the highest value of log $\nu_p$ is 17.2. When the range of log $\nu_p \leq 17$ alone is considered, the two figures become consistent. We note also that Paper I considered BLOs alone, and no redshift correction was applied to their data.

\begin{figure}
\resizebox{\hsize}{!}{\includegraphics{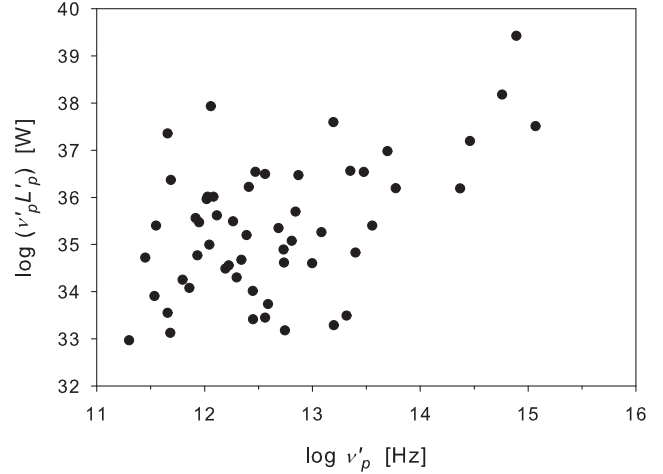}}
\caption{The Doppler-corrected synchrotron peak luminosity vs. peak frequency correlation for bona fide blazars.}
\label{blazars}
\end{figure}

In the bottom panel of Fig.~\ref{lum_vs_peak_z}, we observe the effect of applying the Doppler-correction to the correlation. It is evident that the correlation is altered dramatically. For the entire sample, the results of the Spearman test are $\rho$ = 0.353 and $P <$ 0.001, which implies that the correlation is \textit{positive}. In Table~\ref{spearman_zcorr}, it is evident that, for BLOs, there is a remarkably clear positive correlation between $\nu_p L_p$ and $\nu_p$. The correlations can be interpreted by comparison with Fig.~\ref{Dvar_vs_peak}. The sources of low $\nu_p$ are more boosted, and their luminosity is lowered more significantly by the $D$-correction. For the blazar sample alone (Fig.~\ref{blazars}), we also find a significant positive correlation ($\rho$ = 0.366 and $P$ = 0.003). Thus, we conclude that the blazar sequence, when defined as an anticorrelation of $\nu_p L_p$ and $\nu_p$, is an artefact of variable Doppler boosting across the peak frequency range.

\begin{figure}
\resizebox{\hsize}{!}{\includegraphics{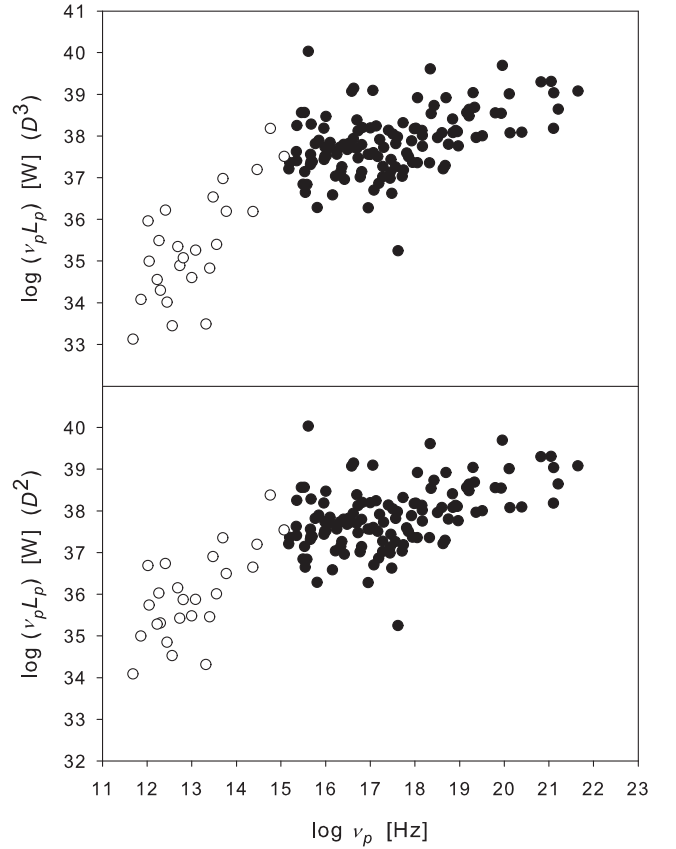}}
\caption{Synchrotron peak luminosity versus peak frequency correlation for BLOs. The data are combined from this work (open circles) and Paper I (filled circles). The former are $D$-corrected, the latter are only $z$-corrected (i.e., we assume that they are not Doppler-boosted). In the top panel, the datapoints represented by open circles are $D^3$ -corrected, in the bottom panel they are $D^2$ -corrected. See text for details.}
\label{BLOs_combined}
\end{figure}

Interestingly, the positive correlation between the $D$-corrected $\nu_p L_p$ and $\nu_p$ for BL Lac objects can explain the peculiar U-shape in the $\nu_p L_p$ versus $\nu_p$ -correlation of Paper I (their Fig.~6). From Fig.~\ref{Dvar_vs_peak}, we deduce that for log $\nu_p \geq$ 15, $D_{var}$ is probably small, and the $D$-correction has little effect (see also Fig.~13 of \citet{wu07}). With this in mind, we considered datapoints with both log $\nu_p \geq$ 15 and a measured redshift from Paper I, and applied a $z$-correction to the $\nu_p$-values. We then plotted these corrected data together with the $D$-corrected datapoints for BLOs derived in this paper, producing the top panel of Fig.~\ref{BLOs_combined}. After combining BLO data sets this way, we are able to observe an indisputable positive correlation ($\rho$ = 0.637 and $P <$ 0.001). It appears that the "intrinsic BLO sequence" implies that sources with higher $\nu_p$ have higher $\nu_p L_p$: this is completely the opposite conclusion to that of the original scenario. A similar trend was observed in two BLOs, 1101+384 (Mrk 421) and 1652+398 (Mrk 501), during flares \citep{pian98,takahashi00}. Their synchrotron peaks were shifted to higher frequencies by the luminosity increase triggered by the flare, in contrast to typical observations of the blazar sequence. For the case of 1652+398 in particular the shift was very pronounced. The positive correlation between peak frequency and luminosity is also in agreement with the number counts of low-energy BLOs (LBLs) and high-energy BLOs (HBLs), which indicate that HBLs are far less numerous than LBLs. As discussed in \citet{padovani07_dxrbs}, this is difficult to understand if HBLs are intrinsically less luminous than LBLs.

As a cross-check, we plotted the BLO correlation using in addition the $D^2$ -correction, that is the exponent $2+\alpha$ in Eq.~\ref{doplum} (bottom panel of Fig.~\ref{BLOs_combined}). Both the content of the figure and the correlation coefficients of the combined BLO sequence changed only marginally. We also checked the $D$ -corrected results of Table~\ref{spearman_zcorr} using the $D^2$ -correction. The significance of the positive correlations for the entire sample and the low-$\nu_p$ BLOs separately persists, while for HPQs, LPQs, and QSOs, $P \geq 0.05$. The main result of this paper, the disappearance of the blazar sequence with Doppler-correction, is therefore unchanged.

\section{Discussion}
\label{dis}

As described in Paper I, the parabolic SED fit can overestimate the synchrotron peak frequency. This occurs in particular when the synchrotron peak occurs at or close to X-ray frequencies, when no datapoints constrain the parabolic fit beyond the X-ray domain. Here we have considered radio-bright quasars and blazars, whose peak frequencies are low; in this case, a parabolic fit is appropriate and negligible overestimation of the peak frequency occurs.

The intrinsic peak frequencies have been studied by \citet{wu07}, who also found a significant negative correlation between $D$ and $\nu'_p$. This independently corroborates the small Doppler factors of high-energy BLOs, and provides support to the concept of a BLO luminosity sequence presented in Fig.~\ref{BLOs_combined}. After all, the log $\nu_p \geq$ 15 -datapoints in their Fig.~13, for which $D \leq$ 5 approximately, are exactly the same ones, taken from Paper I, than those in the high-$\nu_p$ end of our Fig.~\ref{BLOs_combined}. \citet{wu07} also measured a negative correlation between 408 MHz intrinsic luminosity and Doppler-corrected $\nu_p$. In our view this does not, however, correspond to a blazar sequence, which requires an anticorrelation between $\nu_p$ and the luminosity at the synchrotron peak, $\nu_p L_p$, rather than for any particular frequency band. The correlation between radio luminosity and $\nu_p$ can be generated by the shifting of the synchrotron parabola along the log $\nu$ -axis (from low-energy synchrotron source to high-energy synchrotron source), regardless of whether a genuine blazar sequence exists.

A strong positive correlation between $\nu_p L_p$ and $\nu_p$ for BLOs (Fig.~\ref{BLOs_combined}) was unexpected. For these sources, the cooling effects may decrease and intrinsic luminosity increase as $\nu_p$ increases. However, this phenomenon may not be restricted to BLOs: similar correlations may exist for other AGN classes. Unfortunately, to our knowledge there are no sufficiently large SED data sets that would allow us to investigate their respective luminosity sequences as extensively as that of BLOs. Looking at Table~\ref{spearman_zcorr}, we find that HPQs have a marginally significant positive intrinsic correlation, and QSOs are close to the limit. With more datapoints for high values of $\nu_p$ the correlation might also be significant for QSOs. LPQs, however, appear to differ from other classes, exhibiting no intrinsic correlation at all.

Another open question is the Compton dominance of low-$\nu_p$ sources. \citet{fossati98} and \citet{ghisellini98} found that as $\nu_p$ decreased, the inverse Compton component became more luminous compared with its synchrotron counterpart in the SED. The accurate, observational determination of the shape and size of the Compton component for a large sample of objects would require a large amount of data from X-ray to TeV energies. Such a database is unfortunately not yet available for BLOs nor other AGN.

\section{Conclusions}
\label{con}

We have studied the correlation between the synchrotron peak frequency, $\nu_p$, and the synchrotron peak luminosity, $\nu_p L_p$, and the way that this changes when a Doppler-correction is applied to both quantities. We have used an extensive sample of 135 radio-bright AGN, a large part of which forms a complete flux-limited 1.2 Jy northern AGN sample. Our conclusions are as follows:

\begin{enumerate} 
\item There is a clear negative correlation between the Doppler factor, $D$, and $\nu_p$, independent of the method for calculating $D$. This correlation has been tested using Doppler factors calculated from variability data, core versus extended flux, and synchrotron self-Compton flux. Typically, for sources with log $\nu_p \geq$ 15, $D$ is small, while low-$\nu_p$ sources may have values of $D$ as high as 30 and thus be significantly boosted by comparison.
\item The negative correlation between $\nu_p L_p$ and $\nu_p$ at low $\nu_p$, known as the blazar sequence, is not present when the intrinsic, Doppler-corrected values are used. Instead, the correlation becomes \textit{positive}. This is true for the entire AGN sample and a subsample of bona fide blazars. \textit{The blazar sequence is an artefact of variable Doppler boosting across the peak frequency range.}
\item For BLOs in particular, there appears to be a strong intrinsic positive correlation between $\nu_p L_p$ and $\nu_p$. This correlation is further strengthened when the BLO sample in this work is combined with the BLO sample of \citet{nieppola06}. The positive correlation is also in accordance with number counts of LBLs and HBLs.
\end{enumerate} 

\begin{acknowledgements}
We gratefully acknowledge the funding from the Academy of Finland (project numbers 205793, 212656, and 210338). This research has made use of the NASA/IPAC Extragalactic Database (NED) which is operated by the Jet Propulsion Laboratory, California Institute of Technology, under contract with the National Aeronautics and Space Administration. E.N. thanks E.J. Lindfors for help with the high-energy data.
\end{acknowledgements}  


\bibliographystyle{aa}
\bibliography{bibfile}


\longtab{1}{

\begin{longtable}{llccccccc}

\caption{\label{sample}The sample and the observational synchrotron peak frequencies obtained in this work. The values of $\nu_p$ listed here have been determined directly from the SEDs and are not corrected for redshift nor for Doppler boosting.}\\

\hline\hline
Source & Alias & R.A.(J2000) & Dec(J2000) & 1.2 Jy AGN sample & Class & log $\nu_p$ & $z$ & Blazar sample\\
\hline
\endfirsthead
\caption{continued.}\\
\hline\hline
Source & Alias & R.A.(J2000) & Dec(J2000) & 1.2 Jy AGN sample & Class & log $\nu_p$ & $z$ & Blazar sample\\
\hline
\endhead
\hline
\endfoot
0003-066 & NRAO 5 & 00:06:13.90 & -06:23:35.34 &  & BLO & 12.80 & 0.347 & * \\
0007+106 & PG 0007+106 & 00:10:31.01 & +10:58:29.50 & * & GAL & 13.34 & 0.089 &  \\
0014+813 &  & 00:17:08.47 & +81:35:08.10 & * & QSO & 13.91 & 3.366 &  \\
0016+731 &  & 00:19:45.79 & +73:27:30.02 & * & LPQ & 12.82 & 1.781 &  \\
0048-097 & PKS 0048-097 & 00:50:41.32 & -09:29:05.21 &  & BLO & 13.28 & 0.300 & * \\
0059+581 &  & 01:02:45.76 & +58:24:11.14 & * & QSO & 12.49 & 0.644 &  \\
0106+013 & OC 012 & 01:08:38.77 & +01:35:00.30 &  & HPQ & 12.73 & 2.107 &  \\
0109+224 & S2 0109+22 & 01:12:05.80 & +22:44:39.00 & * & BLO & 13.81 &  & * \\
0133+476 & DA 55 & 01:36:58.59 & +47:51:29.10 & * & HPQ & 12.77 & 0.860 & * \\
0149+218 &  & 01:52:18.06 & +22:07:07.70 & * & LPQ & 13.04 & 1.320 &  \\
0202+149 & 4C 15.05 & 02:04:50.41 & +15:14:11.00 & * & HPQ & 12.28 & 0.405 & * \\
0212+735 &  & 02:17:30.81 & +73:49:32.60 & * & HPQ & 12.34 & 2.367 & * \\
0215+015 &  & 02:17:48.95 & +01:44:49.70 & * & HPQ & 13.49 & 1.715 &  \\
0218+357 & S4 0218+357 & 02:21:05.40 & +35:56:15.00 & * & QSO & 12.75 & 0.936 &  \\
0219+428 & 3C 66A & 02:22:39.60 & +43:02:08.00 & * & BLO & 14.96 & 0.444 & * \\
0224+671 &  & 02:28:50.05 & +67:21:03.00 & * & QSO & 12.66 & 0.523 &  \\
0229+131 &  & 02:31:45.89 & +13:22:54.72 & * & LPQ & 13.01 & 2.059 &  \\
0234+285 & 4C 28.07 & 02:37:52.41 & +28:48:09.00 & * & HPQ & 12.81 & 1.207 &  \\
0235+164 & AO 0235+164 & 02:38:38.80 & +16:36:59.00 & * & BLO & 13.31 & 0.940 & * \\
0248+430 &  & 02:51:34.54 & +43:15:15.83 & * & LPQ & 13.45 & 1.311 &  \\
0300+470 & 4C 47.08 & 03:03:35.20 & +47:16:17.00 & * & BLO & 14.34 & 0.475 & * \\
0306+102 & PKS 0306+102 & 03:09:03.60 & +10:29:16.00 & * & BLO & 12.79 & 0.863 & * \\
0316+413 & 3C 84 & 03:19:48.16 & +41:30:42.10 & * & GAL & 14.40 & 0.017 & * \\
0326+349 &  & 03:29:15.35 & +35:10:08.10 & * &  & 12.99 & 0.500 &  \\
0333+321 & NRAO 140 & 03:36:30.11 & +32:18:29.30 & * & LPQ & 13.28 & 1.259 &  \\
0336-019 & CTA 026 & 03:39:30.94 & -01:46:36.00 &  & HPQ & 13.09 & 0.852 & * \\
0355+508 & NRAO 150 & 03:59:29.75 & +50:57:50.20 & * & QSO & 12.53 & 1.510 &  \\
0415+379 & 3C 111 & 04:18:21.28 & +38:01:35.80 & * & GAL & 13.64 & 0.049 &  \\
0420-014 & OA 129 & 04:23:15.80 & -01:20:33.10 &  & HPQ & 12.87 & 0.915 & * \\
0422+004 & OF 038 & 04:24:46.80 & +00:36:07.00 & * & BLO & 14.83 & 0.310 & * \\
0430+052 & 3C 120 & 04:33:11.10 & +05:21:15.60 & * & GAL & $13.95^1$ & 0.033 & * \\
0440-003 & NRAO 190 & 04:42:38.66 & -00:17:43.00 &  & HPQ & 13.50 & 0.844 & * \\
0446+112 & PKS 0446+112 & 04:49:07.67 & +11:21:28.00 & * & GAL & 12.58 & 1.207 &  \\
0454+039 &  & 04:56:47.10 & +04:00:53.00 &  & LPQ & 14.11 & 1.343 &  \\
0458-020 & PKS 0458-020 & 05:01:12.81 & -01:59:14.00 &  & HPQ & 12.62 & 2.286 &  \\
0507+179 &  & 05:10:02.37 & +18:00:41.58 & * & QSO & 12.71 & 0.416 &  \\
0528+134 & PKS 0528+134 & 05:30:56.42 & +13:31:55.15 & * & LPQ & 12.45 & 2.070 & * \\
0552+398 & DA 193 & 05:55:30.81 & +39:48:49.17 & * & LPQ & 12.12 & 2.365 &  \\
0605-085 & PKS 0605-085 & 06:07:59.70 & -08:34:50.00 &  & HPQ & 13.40 & 0.872 & * \\
0642+449 & OH 471 & 06:46:32.03 & +44:51:16.59 & * & LPQ & 12.39 & 3.406 &  \\
0716+714 & S5 0716+714 & 07:21:53.30 & +71:20:36.00 & * & BLO & 13.87 & 0.300 & * \\
0723-008 & PKS 0723-008 & 07:25:50.64 & +00:54:56.54 &  & GAL & 14.39 & 0.127 &  \\
0735+178 & PKS 0735+17 & 07:38:07.40 & +17:42:19.00 & * & BLO & 14.07 & 0.424 & * \\
0736+017 &  & 07:39:18.03 & +01:37:04.60 & * & HPQ & 13.42 & 0.191 & * \\
0738+313 &  & 07:41:10.70 & +31:12:00.23 & * & GAL & 13.54 & 0.631 &  \\
0748+126 &  & 07:50:52.05 & +12:31:04.83 & * & LPQ & 12.76 & 0.889 &  \\
0754+100 & PKS 0754+100 & 07:57:06.64 & +09:56:34.90 & * & BLO & 13.70 & 0.266 & * \\
0804+499 &  & 08:08:39.67 & +49:50:36.50 & * & HPQ & 12.75 & 1.430 &  \\
0814+425 & OJ 425 & 08:18:16.00 & +42:22:45.40 & * & BLO & 13.26 & 0.258 & * \\
0823+033 & PKS 0823+033 & 08:25:50.30 & +03:09:24.00 & * & HPQ & 12.73 & 0.506 & * \\
0827+243 & OJ 248 & 08:30:52.08 & +24:10:60.00 & * & LPQ & 12.94 & 0.941 &  \\
0829+046 & PKS 0829+046 & 08:31:48.88 & +04:29:39.09 & * & BLO & 13.92 & 0.180 & * \\
0836+710 & 4C 71.07 & 08:41:24.37 & +70:53:42.20 & * & LPQ & 13.94 & 2.170 & * \\
0846+513 &  & 08:49:57.98 & +51:08:29.00 &  & QSO & 13.24 & 1.860 &  \\
0850-1213 &  & 08:50:09.60 & -12:13:34.00 &  & QSO & 12.58 & 0.566 &  \\
0851+202 & OJ 287 & 08:54:48.80 & +20:06:30.00 & * & BLO & 13.64 & 0.306 & * \\
0906+430 & 3C 216 & 09:09:33.50 & +42:53:46.08 & * & HPQ & 12.99 & 0.670 & * \\
0917+449 &  & 09:20:58.46 & +44:41:53.99 & * & QSO & 12.85 & 2.180 &  \\
0923+392 & 4C 39.25 & 09:27:03.01 & +39:02:20.90 & * & LPQ & 11.99 & 0.699 & * \\
0945+408 & 4C 40.24 & 09:48:55.34 & +40:39:44.60 & * & LPQ & 12.87 & 1.252 &  \\
0953+254 &  & 09:56:49.88 & +25:15:16.05 & * & LPQ & 13.03 & 0.712 &  \\
0954+556 & S4 0954+556 & 09:57:38.24 & +55:22:58.00 & * & HPQ & 13.77 & 0.901 & * \\
0954+658 & S4 0954+65 & 09:58:47.20 & +65:33:54.00 & * & BLO & 14.16 & 0.367 & * \\
1019+222 &  & 10:21:54.59 & +21:59:31.20 & * & GAL &  & 1.617 &  \\
1036+054 &  & 10:38:46.78 & +05:12:29.09 & * &  & 12.57 & 0.473 &  \\
1040+244 & TEX 1040+244 & 10:43:09.00 & +24:08:35.00 & * & BLO & 12.82 & 0.560 & * \\
1039+811 &  & 10:44:23.06 & +80:54:39.44 & * & LPQ & 13.04 & 1.260 &  \\
1049+215 &  & 10:51:48.79 & +21:19:52.31 & * & LPQ & 12.84 & 1.300 &  \\
1055+018 & OL 093 & 10:58:29.61 & +01:33:58.80 & * & HPQ & 12.68 & 0.888 & * \\
1150+812 &  & 11:53:12.50 & +80:58:29.15 & * & QSO & 12.75 & 1.250 &  \\
1150+497 &  & 11:53:24.47 & +49:31:08.83 & * & HPQ & 13.29 & 0.334 &  \\
1156+295 & 4C 29.45 & 11:59:31.83 & +29:14:44.00 & * & HPQ & 13.44 & 0.729 &  \\
1219+285 & ON 231 & 12:21:31.70 & +28:13:58.00 & * & BLO & 15.10 & 0.102 & * \\
1219+044 &  & 12:22:22.55 & +04:13:15.78 & * & QSO & 13.07 & 0.965 &  \\
1222+216 & PKS 1222+216 & 12:24:54.51 & +21:22:47.00 & * & QSO & 13.27 & 0.435 &  \\
1226+023 & 3C 273 & 12:29:06.69 & +02:03:08.60 & * & LPQ & 13.89 & 0.158 & * \\
1253-055 & 3C 279 & 12:56:11.17 & -05:47:21.50 &  & HPQ & 12.67 & 0.538 & * \\
1308+326 & AUCVn & 13:10:28.66 & +32:20:43.80 & * & BLO & 13.49 & 0.992 & * \\
1324+224 &  & 13:27:00.86 & +22:10:50.16 & * & QSO & 12.74 & 1.400 &  \\
1406-076 & PKS 1406-076 & 14:08:56.48 & -07:52:27.00 &  & QSO & 12.69 & 1.494 &  \\
1413+135 & PKS 1413+135 & 14:15:58.80 & +13:20:24.00 & * & BLO & 12.64 & 0.247 & * \\
1418+546 & OQ 530 & 14:19:46.60 & +54:23:14.00 & * & BLO & 14.03 & 0.152 & * \\
1502+106 & OR 103 & 15:04:24.98 & +10:29:39.00 & * & HPQ & 13.08 & 1.833 & * \\
1510-089 & PKS 1510-089 & 15:12:50.53 & -09:05:59.00 &  & HPQ & 13.49 & 0.361 & * \\
1532+016 & PKS 1532+016 & 15:34:52.45 & +01:31:04.00 & * & HPQ & 13.23 & 1.435 &  \\
1538+149 & 4C 14.60 & 15:40:46.50 & +14:47:45.90 & * & BLO & 13.84 & 0.605 & * \\
1546+027 &  & 15:49:29.44 & +02:37:01.20 & * & HPQ & 12.97 & 0.413 &  \\
1548+056 &  & 15:50:35.27 & +05:27:10.45 & * & HPQ & 12.74 & 1.422 &  \\
1606+106 & 4C 10.45 & 16:08:46.20 & +10:29:07.80 & * & LPQ & 12.80 & 1.226 &  \\
1611+343 & DA 406 & 16:13:41.00 & +34:12:48.00 & * & LPQ & 12.69 & 1.401 & * \\
1633+382 & 4C 38.41 & 16:35:15.49 & +38:08:04.50 & * & LPQ & 12.89 & 1.814 & * \\
1637+574 & OS 562 & 16:38:13.46 & +57:20:24.00 & * & LPQ & 13.17 & 0.745 & * \\
1642+690 &  & 16:42:07.85 & +68:56:39.76 & * & HPQ & 12.54 & 0.751 &  \\
1641+399 & 3C 345 & 16:42:58.81 & +39:48:37.00 & * & HPQ & 12.71 & 0.595 & * \\
1652+398 & MRK 501 & 16:53:52.20 & +39:45:36.00 & * & BLO & 17.19 & 0.034 & * \\
1725+044 & PKS 1725+044 & 17:28:24.95 & +04:27:05.00 & * & QSO & 13.27 & 0.297 & * \\
1730-130 & NRAO 530 & 17:33:02.71 & -13:04:49.55 &  & QSO & 12.42 & 0.902 &  \\
1739+522 & S4 1739+52 & 17:40:36.98 & +52:11:43.00 & * & HPQ & 12.93 & 1.379 &  \\
1741-038 & PKS 1741-038 & 17:43:58.86 & -03:50:04.60 &  & HPQ & 12.35 & 1.054 & * \\
1749+096 & PKS 1749+096 & 17:51:32.70 & +09:39:01.00 & * & BLO & 12.78 & 0.320 & * \\
1758+388 &  & 18:00:24.72 & +38:48:31.10 & * & QSO & 12.70 & 2.092 &  \\
1803+784 & S5 1803+784 & 18:00:45.40 & +78:28:04.00 & * & BLO & 13.60 & 0.684 & * \\
1807+698 & 3C 371 & 18:06:50.70 & +69:49:28.00 & * & BLO & 14.46 & 0.051 & * \\
1823+568 & 4C 56.27 & 18:24:07.07 & +56:51:01.50 & * & BLO & 12.93 & 0.663 & * \\
1828+487 & 3C 380 & 18:29:31.80 & +48:44:46.62 & * & LPQ & 13.88 & 0.692 &  \\
1845+797 & 3C 390.3 & 18:42:08.99 & +79:46:17.00 & * & GAL &  & 0.055 &  \\
1901+319 &  & 19:02:55.93 & +31:59:41.70 & * & QSO & 15.67 & 0.635 &  \\
1926+61 & S4 1926+61 & 19:27:30.40 & +61:17:32.00 & * & BLO & 13.27 &  & * \\
1928+738 & 4C 73.18 & 19:27:48.50 & +73:58:01.60 & * & LPQ & 13.32 & 0.303 & * \\
1954+513 &  & 19:55:42.74 & +51:31:48.55 & * & LPQ & 12.90 & 1.223 &  \\
2007+776 & S5 2007+77 & 20:05:31.10 & +77:52:43.00 & * & BLO & 12.95 & 0.342 & * \\
2005+403 &  & 20:07:44.94 & +40:29:48.60 & * & QSO & 13.07 & 1.736 &  \\
2021+614 & OW 637 & 20:22:06.68 & +61:36:58.80 & * & LPQ & 12.67 & 0.227 & * \\
2022+171 &  & 20:24:56.56 & +17:18:13.20 &  & LPQ & 13.14 & 1.050 &  \\
2037+511 &  & 20:38:37.04 & +51:19:12.66 & * & QSO & 12.26 & 1.687 &  \\
2059+034 &  & 21:01:38.83 & +03:41:31.30 & * & QSO & 13.03 & 1.013 &  \\
2121+053 &  & 21:23:44.52 & +05:35:22.09 & * & HPQ & 13.01 & 1.941 &  \\
2134+004 & OX 057 & 21:36:38.59 & +00:41:54.21 & * & LPQ & 12.32 & 1.936 & * \\
2136+141 &  & 21:39:01.31 & +14:23:36.00 & * & LPQ & 12.71 & 2.427 &  \\
2141+175 &  & 21:43:35.55 & +17:43:48.50 & * & LPQ & 13.70 & 0.211 &  \\
2144+092 &  & 21:47:10.16 & +09:29:46.67 &  & QSO & 13.04 & 1.113 &  \\
2145+067 &  & 21:48:05.46 & +06:57:38.60 & * & LPQ & 12.73 & 0.990 & * \\
2200+420 & BL LAC & 22:02:43.30 & +42:16:39.00 & * & BLO & 14.14 & 0.070 & * \\
2201+315 & 4C 31.63 & 22:03:14.98 & +31:45:38.30 & * & LPQ & 13.63 & 0.298 &  \\
2201+171 &  & 22:03:26.89 & +17:25:48.20 & * & QSO & 12.90 & 1.075 &  \\
2209+236 & PKS 2209+236 & 22:12:05.97 & +23:55:40.00 & * & QSO & 12.62 & 1.125 &  \\
2223-052 & 3C 446 & 22:25:47.30 & -04:57:01.39 &  & BLO & 12.74 & 1.404 & * \\
2227-088 &  & 22:29:40.09 & -08:32:54.50 &  & HPQ & 12.99 & 1.562 &  \\
2230+114 & CTA 102 & 22:32:36.41 & +11:43:50.90 & * & HPQ & 13.19 & 1.037 & * \\
2234+282 &  & 22:36:22.47 & +28:28:57.41 & * & HPQ & 12.74 & 0.795 &  \\
2251+158 & 3C 454.3 & 22:53:57.75 & +16:08:53.60 & * & HPQ & 12.83 & 0.859 & * \\
2254+074 & PKS 2254+074 & 22:57:17.30 & +07:43:12.27 &  & BLO & 14.07 & 0.190 & * \\
2344+092 &  & 23:46:36.84 & +09:30:45.52 & * & LPQ & 14.25 & 0.677 & * \\
2351+456 & 4C 45.51 & 23:54:21.68 & +45:53:04.00 & * & QSO & 12.44 & 1.986 &  \\
2353+816 &  & 23:56:22.79 & +81:52:52.26 & * & QSO & 12.74 & 1.344 &  \\
\hline
\end{longtable}
\begin{list}{}{\setlength{\leftmargin}{12pt}}
\item \footnotesize{$^{\mathrm{1}}$ = log $\nu_p$ determined from the highest flux density measurement of the synchrotron component of the SED (see Sect.~\ref{seds}).}
\end{list}
}



\Online

\begin{figure*}
\resizebox{\hsize}{!}{\includegraphics{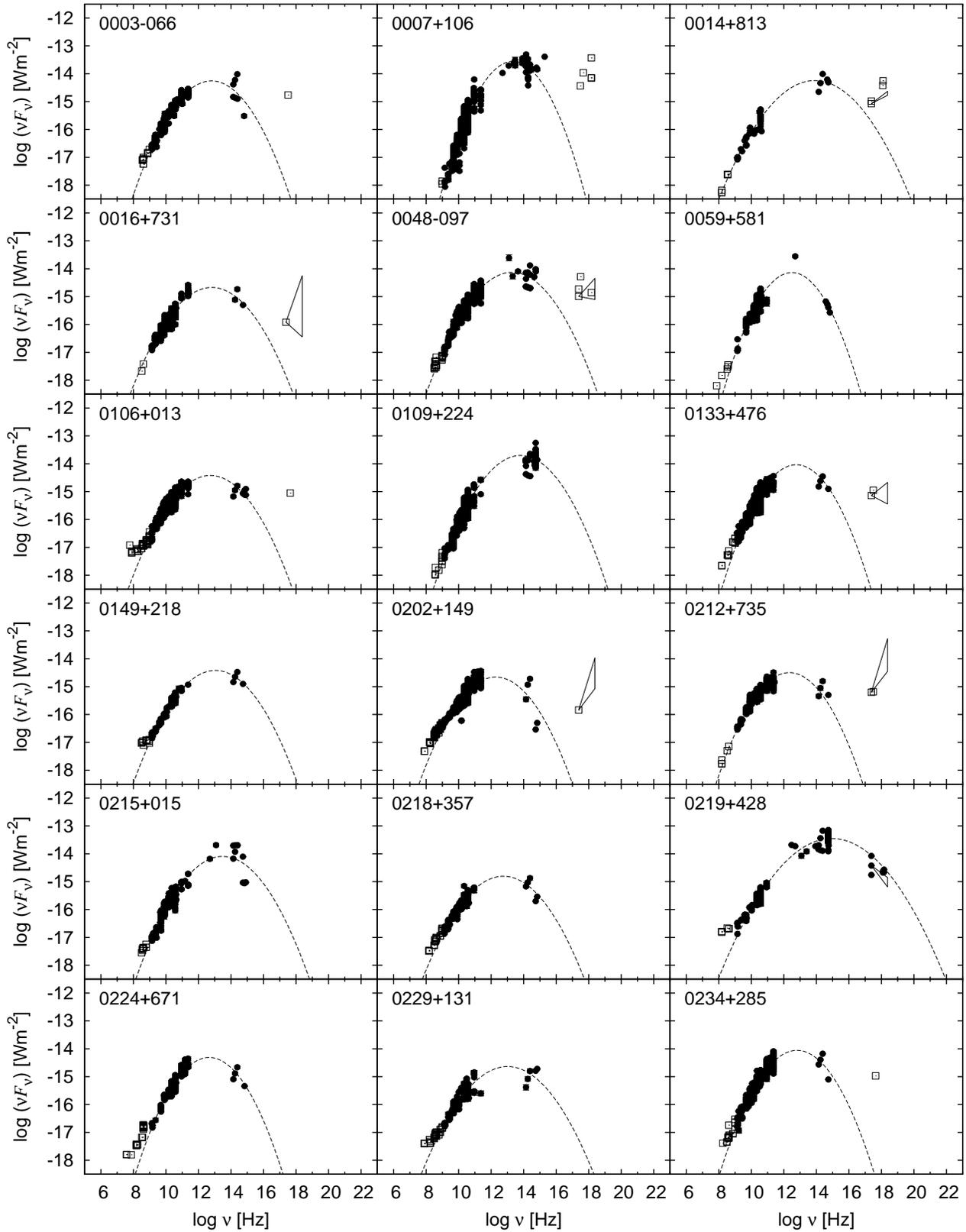}}
\caption{The spectral energy distributions determined in this work. All quantities are observational and uncorrected for redshift and Doppler boosting. The dashed line represents the parabolic fit to the synchrotron component. Datapoints marked with filled circles are included in the fit, the empty squares are not. Sources 1019+222 and 1845+797 could not be fitted with a parabolic function.}
\label{sedfig}
\end{figure*}

\addtocounter{figure}{-1}
\begin{figure*}
\resizebox{\hsize}{!}{\includegraphics{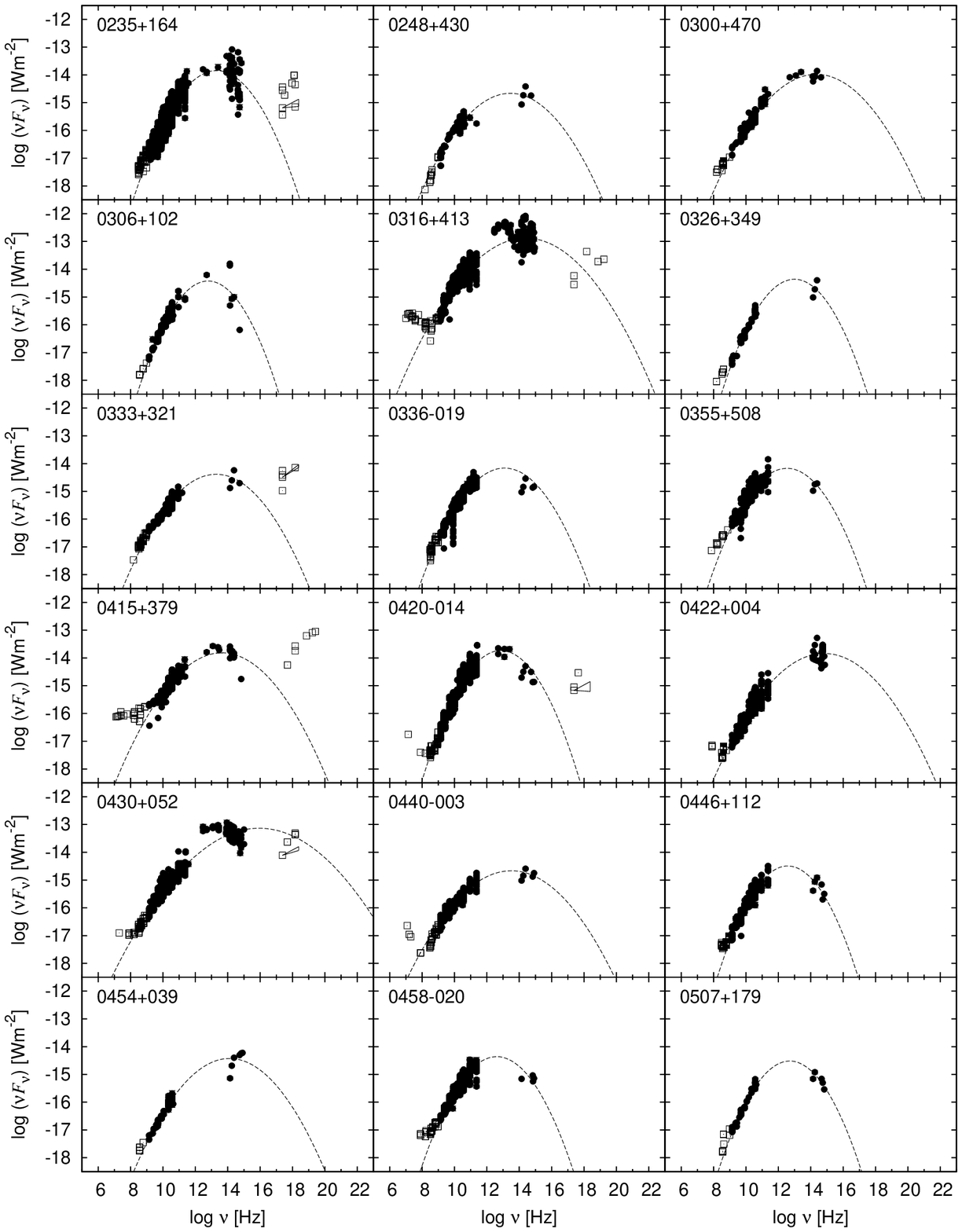}}
\caption{continued.}
\end{figure*}

\addtocounter{figure}{-1}
\begin{figure*}
\resizebox{\hsize}{!}{\includegraphics{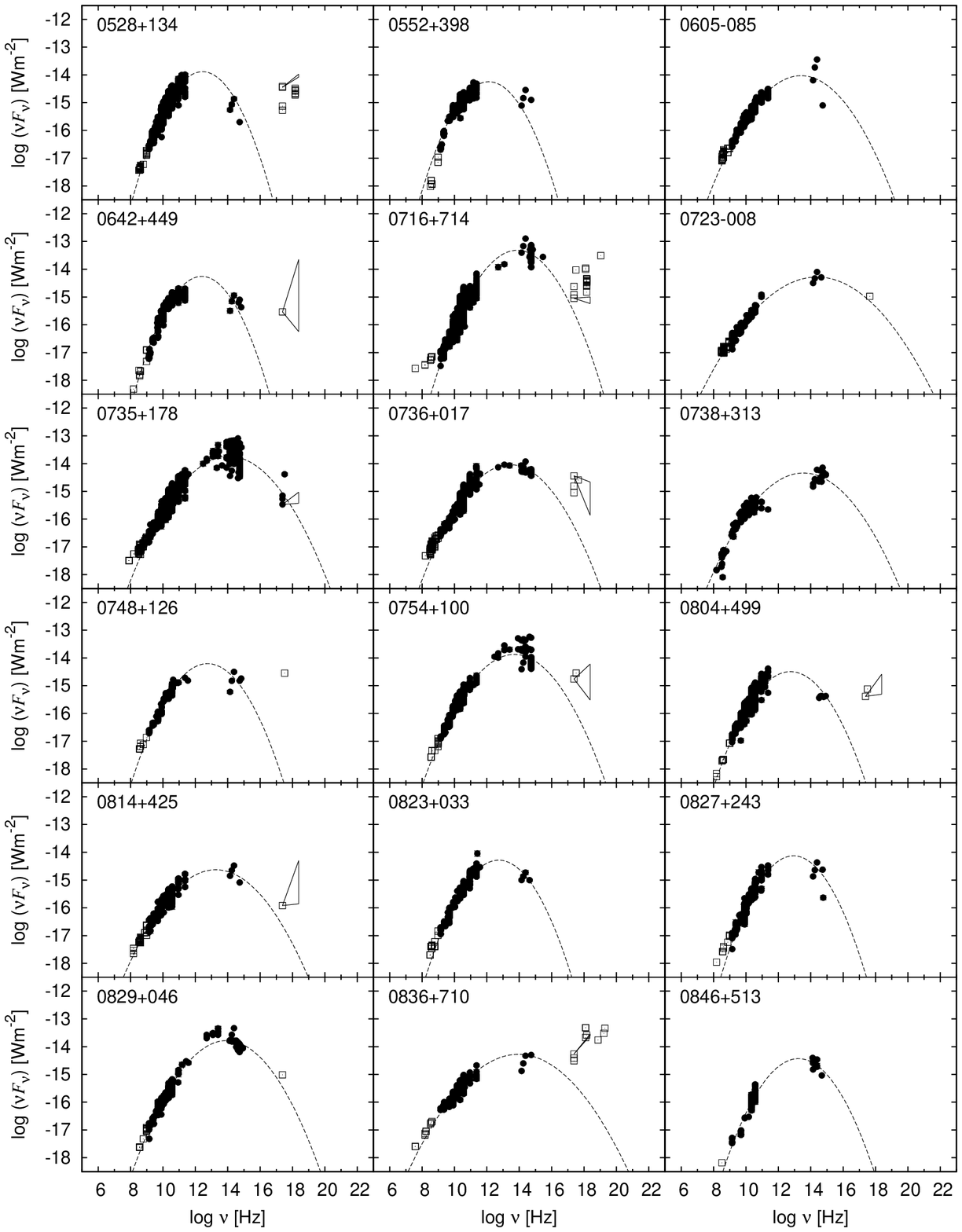}}
\caption{continued.}
\end{figure*}

\addtocounter{figure}{-1}
\begin{figure*}
\resizebox{\hsize}{!}{\includegraphics{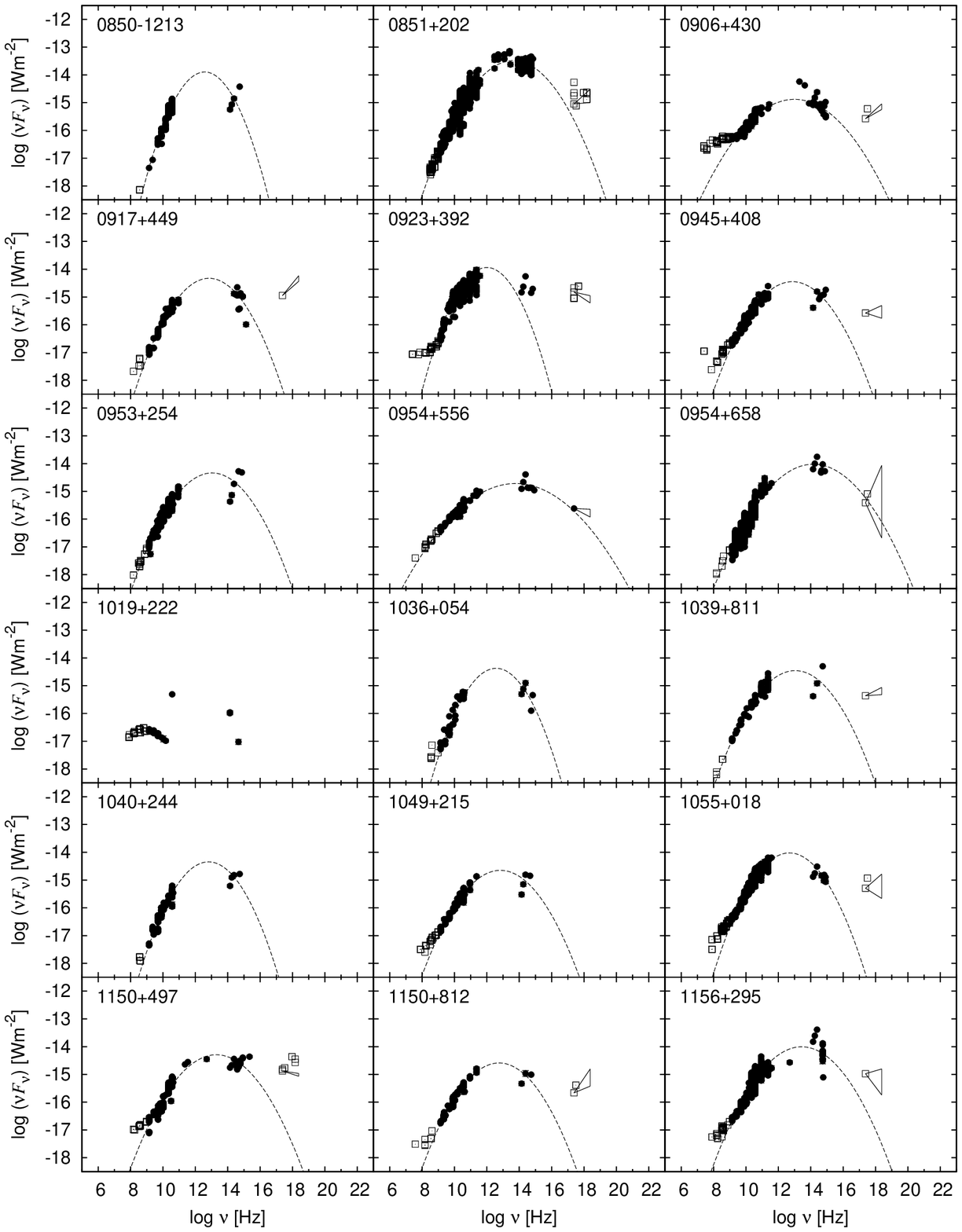}}
\caption{continued.}
\end{figure*}

\addtocounter{figure}{-1}
\begin{figure*}
\resizebox{\hsize}{!}{\includegraphics{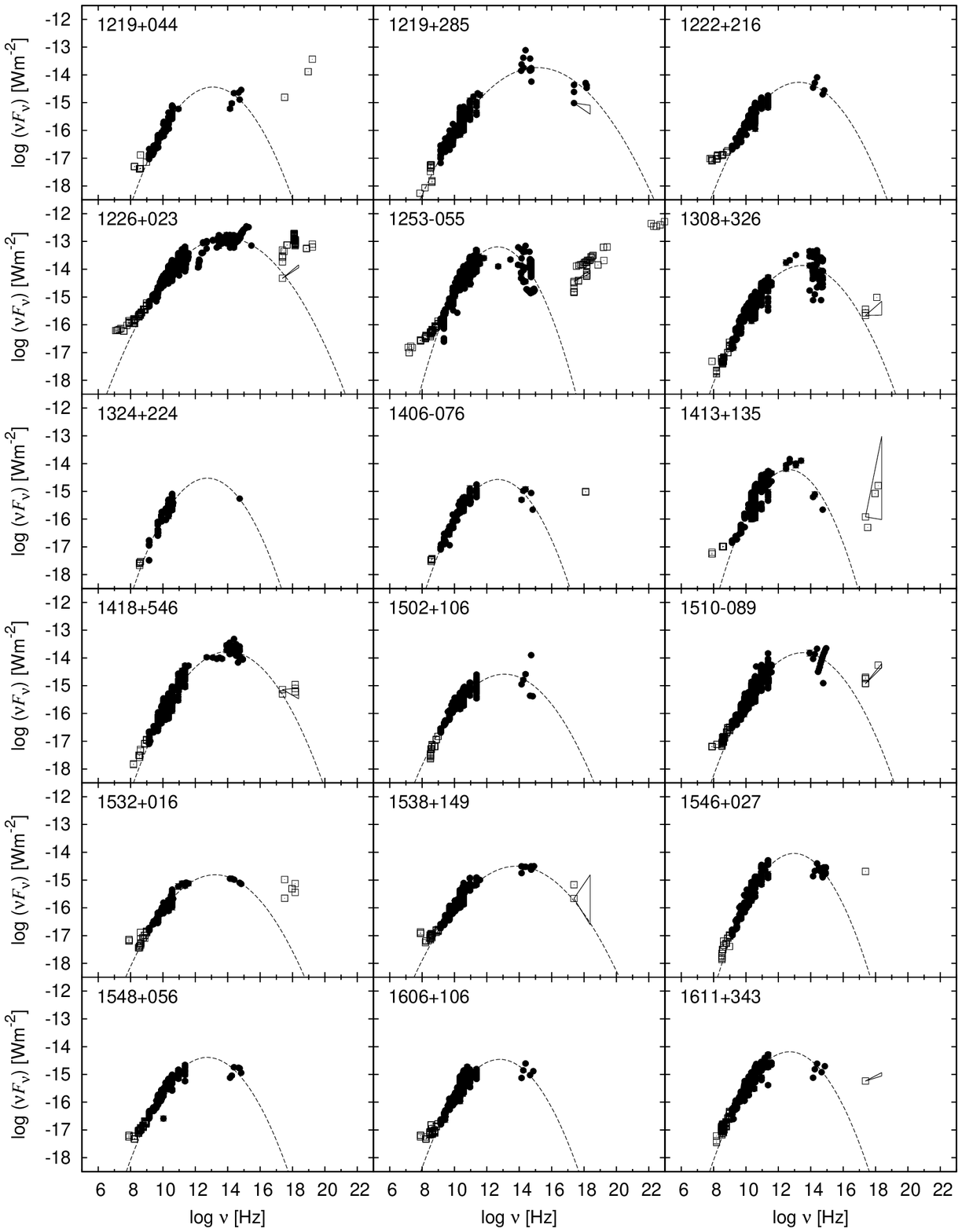}}
\caption{continued.}
\end{figure*}

\addtocounter{figure}{-1}
\begin{figure*}
\resizebox{\hsize}{!}{\includegraphics{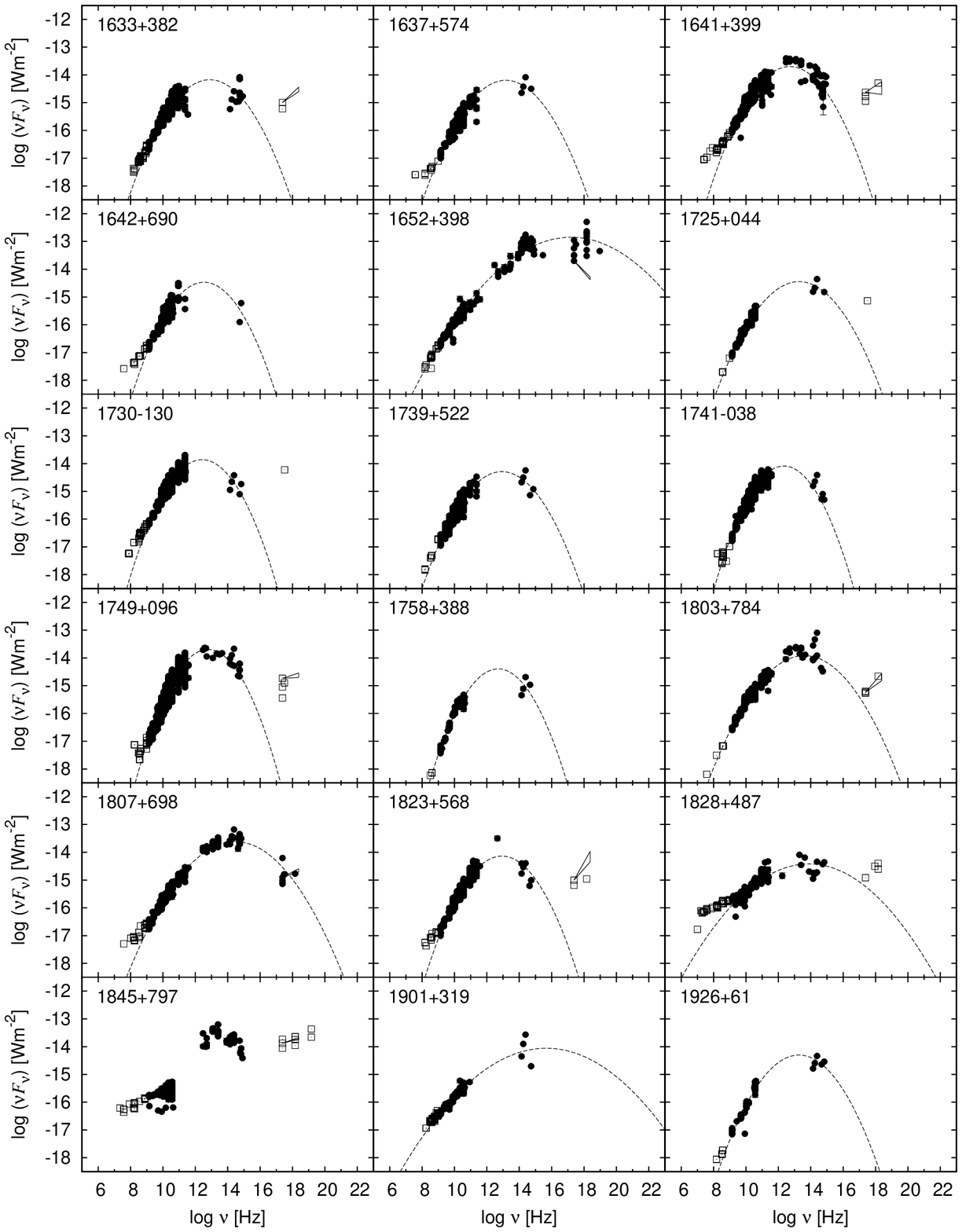}}
\caption{continued.}
\end{figure*}

\addtocounter{figure}{-1}
\begin{figure*}
\resizebox{\hsize}{!}{\includegraphics{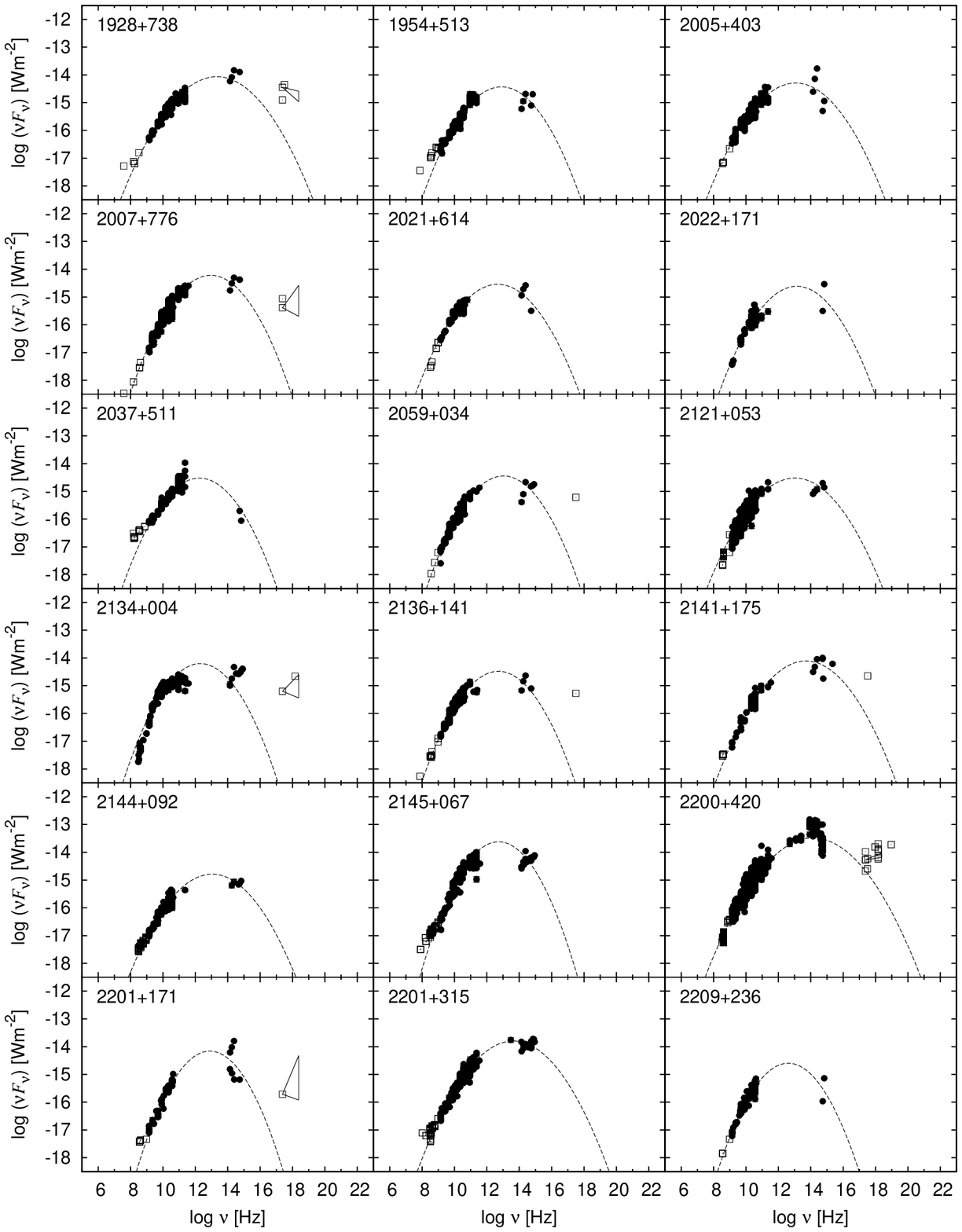}}
\caption{continued.}
\end{figure*}

\addtocounter{figure}{-1}
\begin{figure*}
\resizebox{\hsize}{!}{\includegraphics{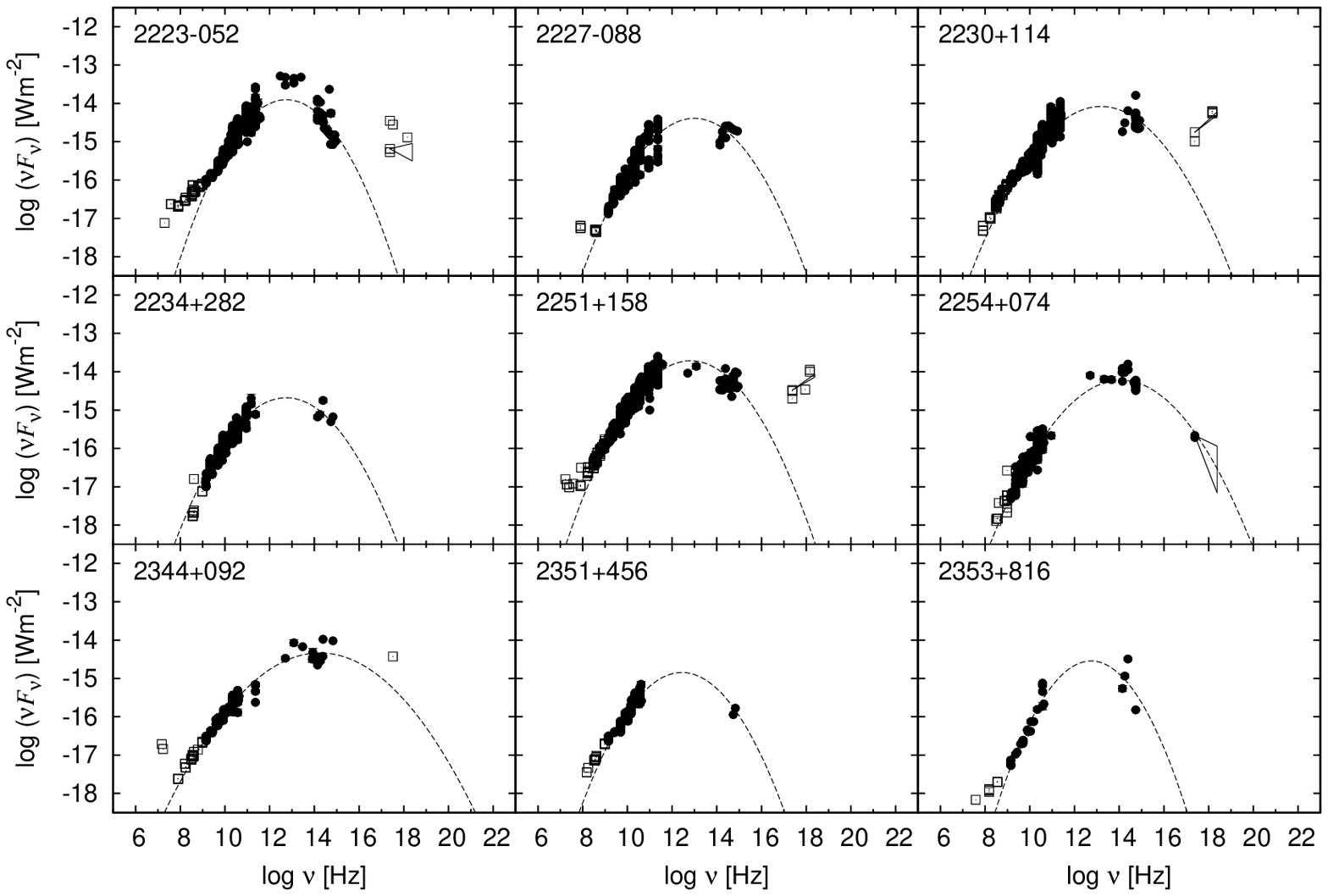}}
\caption{continued.}
\end{figure*}

\end{document}